# Real–time Monitoring and Forecasting of Ecological Processes

**J. Culiţă, A. Dumitraşcu, D. Ştefănoiu**

*Automatic Control and Computers Faculty, "Politehnica" University of Bucharest, Romania
(e-mails: jculita@yahoo.com, dumalex@ecosys.pub.ro, danny@indinf.pub.ro)*

**Abstract**: The paper introduces a real-time monitoring and forecasting system for ecological phenomena. The process yields a collection of ecological parameters viewed as distributed time series, which are measured by means of wireless network of sensors. The acquired data are preliminary processed and modeled by using complex algorithms in view of prediction. There are three graphical user interfaces implemented within the monitoring and forecasting system: eko-View and eko-Greenhouse (which directly interacts with the process) and eko-Forecast (which estimates the future evolution of some ecological parameters). The monitoring system was effectively integrated in an industrial application dealing with automatic irrigation of a small greenhouse. The forecasting simulation results with real data and a comparative assessment of predictor performances are presented in the end.

## 1. INTRODUCTION

Rapid climate changes and the negative impact of industry upon the environment require designing and employing of an automatic monitoring system of geographical areas. The general purpose of monitoring is to forecast the behavior of the ecological system in view of disaster anticipation or avoidance.

The ecological phenomena could be evidenced either in an open space or in an enclosed space. Phenomena like correlation between temperature variation and humidity or heat and humidity transfer usually occur in a greenhouse. Especially in a microclimate, ambient temperature and humidity, dew point and solar radiation are quite correlated, which could improve their prediction accuracy. The soil parameters (moisture, temperature, water content, leaf wetness) are however less correlated.

The paper mainly presents an ecological monitoring and forecasting system *EcoMonFor*, which allows monitoring and forecasting of multi-variable ecological signals both in local and wider geographical regions. EcoMonFor was successfully integrated in a new application on remote monitoring and control of a small greenhouse (Dumitrascu, 2010). Basically, the application aim is the automatic watering of plants, when the ecological state requires it, in order to yield suitable growth of plants. The distributed monitoring and control architecture of the ecological process interconnects several subsystems (see figure 1). The first one is a wireless acquisition and monitoring subsystem structured on three hierarchical levels (Culita and Stefanoiu, 2010) and provided with three graphical user-friendly interfaces eKo-View, eko-Greenhouse and eko-Forecast. The second one is the automatic control subsystem made of PLCs and industrial communication networks. Finally, the irrigation subsystem consists of two water tanks, sensors and actuators.

This article is not pointing to control design. Data monitoring and preparing in view of prediction, together with forecasting results are discussed here only. In our approach, the ecological signal prediction relies on numerical models that were previously implemented as FORWAVER, PARMA, PARMAX, KARMA predictors (Stefanoiu et al, 2008; Stefanoiu and Culita, 2010). It is expected that the forecasting experimental results to be quite accurate especially when the ecological parameter supplied by the greenhouse are correlated to each other.

The paper is structured as follows. Section 2 introduces the distributed architecture for monitoring and control of the greenhouse. Section 3 presents the acquisition and preliminary processing of the ecological parameters provided by the greenhouse. The performances of prediction are shown within Section 4. A conclusion and the references list complete the article.

## 2. MONITORING AND CONTROL SYSTEM ARCHITECTURE OF THE GREENHOUSE

The greenhouse consists of six plants which are located in two separated laboratory rooms in order to create different microclimates. The improper care of plants led to constructing an automatic irrigation system. Figure 1 depicts the distributed monitoring and control architecture of the greenhouse, which integrates: the automatic control system of irrigation (left side down), the irrigation system (left side up) and, most concerned, the ecological monitoring system EcoMonFor (right side).

Constructively, EcoMonFor was separated in two components: a mobile part, referred to as *EcoMonFor-M*, figured inside the (red) ellipsis and a fixed part, namely *EcoMonFor-F*, represented in the right down corner. The mobile monitoring system is structured on three hierarchical levels as shown in the right upper side of figure 1:

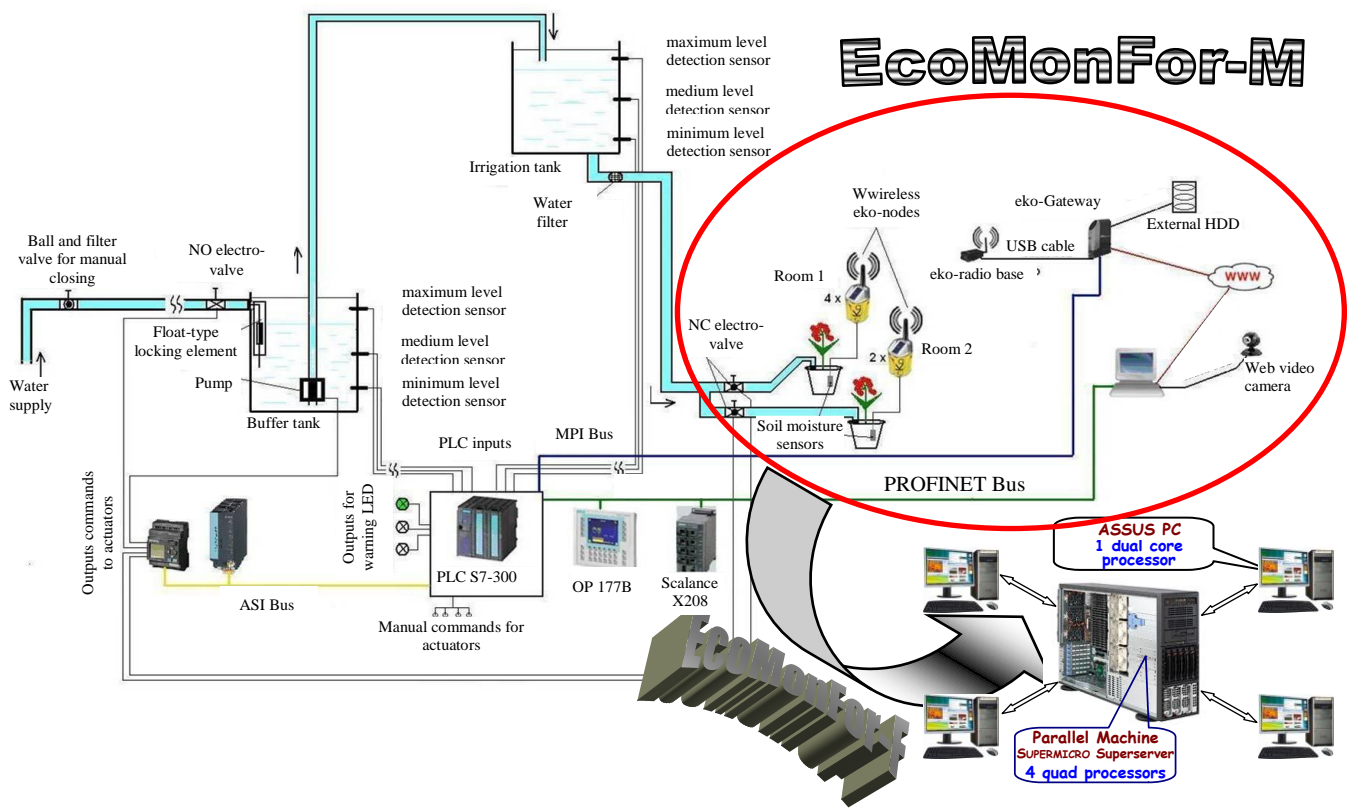

Fig. 1. Architecture of the small greenhouse control system including *EcoMonFor*

the set of wireless eko-sensors; eko-Gateway – a centralizing (kernel) equipment of the sensor network; a computer (or laptop). The last two components are wirelessly connected to Internet, in order to enable running remote applications. Moreover, the computer fulfils the function of real-time video supervision of the whole system through a coupled webcam. EcoMonFor-M mostly is responsible for remote data acquisition and monitoring, which means it could cover an extended geographical area. It can be employed for a quick prediction of the measured data, as well. The data collection supplied by eko-Gateway module is sent to EcoMonFor-F with the aim of high quality prediction of the ecological phenomena. This strategy is suggested by the curved arrow in the bottom of the image. The core of the fixed component consists in a parallel computer with 16 processors. This is connected via internet to an extensible computer network. The central unit is hosting complex algorithms for modeling, identification and forecasting of distributed ecological signals: PARMA, PARMAX, KARMA and FORWAVER.

Both components of EcoMonFor are working on the following strategy: first, the acquisition and the preliminary processing of data are performed. Sometimes, data provided by sensors are damaged and need to be enhanced. Some operations are necessary to improve data (as shown within the next section). The visual monitoring of the greenhouse stands for the second step, which is executed in parallel with the acquisition and is remotely performed by eko-Gateway. Two web graphical user interfaces are implemented on the computer connected to eko-Gateway. The first one, eko-View, gives the user the ability to set and display the configuration of the sensor network and then to start monitoring and acquisition, from anywhere in the world. Moreover, it is supplied with several facilities in handling data (i.e. graphical display of interest data, exporting to common programming environments, setting alerting rules). The sensor configuration on the real case study will be exemplified in the next section. The second web interface is eKo-Greenhouse, as displayed by figure 2. This is more oriented towards the irrigation application. Thus, its role is helping the user to directly and remotely interact with the greenhouse, via internet, by accessing the process parameters and controlling the automatic irrigation system. Technically, the main panel is based on Apache http server and it is password protected. It was built using common Web technologies: HTML, JavaScript, XML PHP. The interface configuration displays four interesting zones: on the left side above, the visual image of the process is permanently offered by a webcam; beneath, the results of the last 10 commands to the actuators are completely shown; at right, four selection buttons are depicted for choosing a corresponding control panel (i.e, commands to the control device PLC S7-300; remote commands for manual control of actuators in the irrigation process and information about the current status of them; displaying and setting the ecological parameters by the user; exporting data from eko-Gateway in a comprehensible and useful format and saving them on external disk, for subsequent processing).

The final step of the operating strategy of EcoMonFor system corresponds to data modeling on prediction purpose. A convivial graphical interface *eKo-Forecast* was implemented in MATLAB programming language, in order to complete a forecasting experiment (Culita and Stefanoiu, 2010).

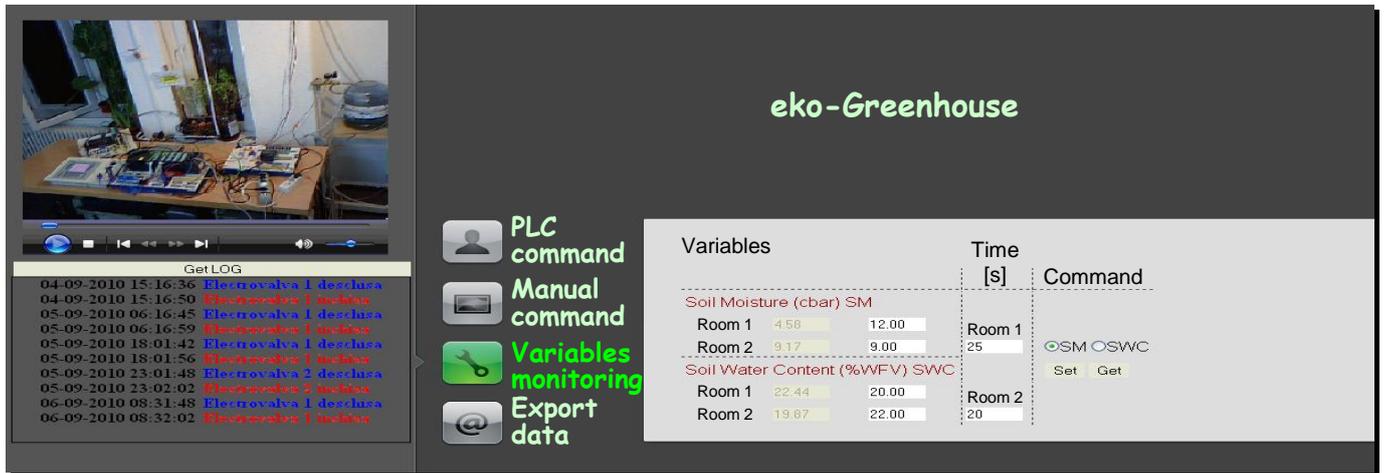

Fig. 2. The web interface eko-Greenhouse, yielding the remote control

It facilitates running PARMA, PARMAX, KARMA and FORWAVER predictors within *FORTIS* (*FOR*ecasting of *TI*me *Series*) simulator. The interface offers a graphical illustration of the forecasting results. Although all predictors can proceed on the same fixed or mobile component either, the faster predictors of FORTIS (PARMA and FORWAVER) are commonly hosted by the mobile entity and the slower algorithms, PARMAX and KARMA, are usually executed on the fixed component.

EcoMonFor system represents an additional part of the irrigation application. One hand, it decides the irrigation commands by acting through the sensor subsystem. On the other hand it processes the measured data to forecast them.

### 3. DATA ACQUISITION AND PRELIMINARY PROCESSING

As mentioned before, the greenhouse consists of 6 plants, which are located in two different rooms. Each plant was associated to a wireless node for acquisition and monitoring purpose. The monitoring can be carried out by using eko-View and eko-Greenhouse interfaces. Figure 3 illustrates the main panel of eko-View interface. The plants are represented by their photos. Every node is capable of transmitting data from at most 4 sensors, while a sensor can measure 1-3 ecological parameters simultaneously (for example, there is a singleton sensor for soil moisture and humidity or for ambient humidity, temperature and dew point). Though, the number of the ecological parameters differs for each sensor.

Figure 3 also shows the synoptic map of the monitored ecological parameters associated to each acquisition node. There have been used 21 sensors, which are transmitting data for 33 ecological parameters, as it can be noticed from the figure. As a major aim of monitoring, we are interested in forecasting some ecological parameters of the greenhouse and validating (testing) the correlations between them. In order to send data to FORTIS simulator (in view of prediction), the parameter values (of the same node) have to be grouped in data blocks, according to their possible correlations. For example, humidity is correlated to temperature which, in its turn, is correlated to solar radiation. It is rather difficult to presume that the soil parameters coming from different plants are correlated each other, taking into account that the plants lie in different locations. Each block corresponds to a node and contains data from 3-4 acquisition channels. The name of such data block is an identification code including: node identity (1-6); parameter type (soil or ambient); the acronyms of the measured parameters. For further processing (modeling and forecasting), the data blocks need to be converted in structures accepted by MATLAB programming environment.

Figure 3, as well as table 1, indicates all the observed parameters of the small greenhouse. One can also see their varying range and measurement units in table 1. Parameters acronyms were set for data indexing and identification purposes.

**Table 1. Ecological parameters of sensors network.**

| Soil | Leaves | Ambient |
|---|---|---|
| Moisture (Mo) 0 ... 240 [cbar] | Leaf Wetness (LeWe) 0 ... 1024 [CntS] | Humidity (Hu) 0 ... 100 [%] |
| Temperature (Te) –40 ... +65 [°C] | | Temperature (Te) –40 ... +65 [°C] |
| Water Content (WaCo) 0 ... 100 [%wfv] | | Dew Point (DwPo) –10 ... 50 [°C] |
| | | Solar Radiation (SoRa) 0 ... 1800 [W/m$^2$] |

The ecological sensors usually provide unsynchronized or faulty data. Therefore, preliminary data processing is necessary. A simple and intuitive method of obtaining synchronized data is the hourly averaging technique. Frequently, there could be missing data on different acquisition channels at some instants. In this case, the interpolation followed by re-sampling can return correct data. First, for isolated missing information, linear interpolation is enough, as it can be noticed from inspecting figure 4 (raw data) and figure 5 (linear interpolated data). Next, for the other lost data (that extends over an interval of sampling instants), autoregressive interpolation (AR) seems to be quite adequate (see results in figure 6). The AR model was identified by applying Levinson-Durbin Algorithms (Soderstrom and Stoica, 1989).

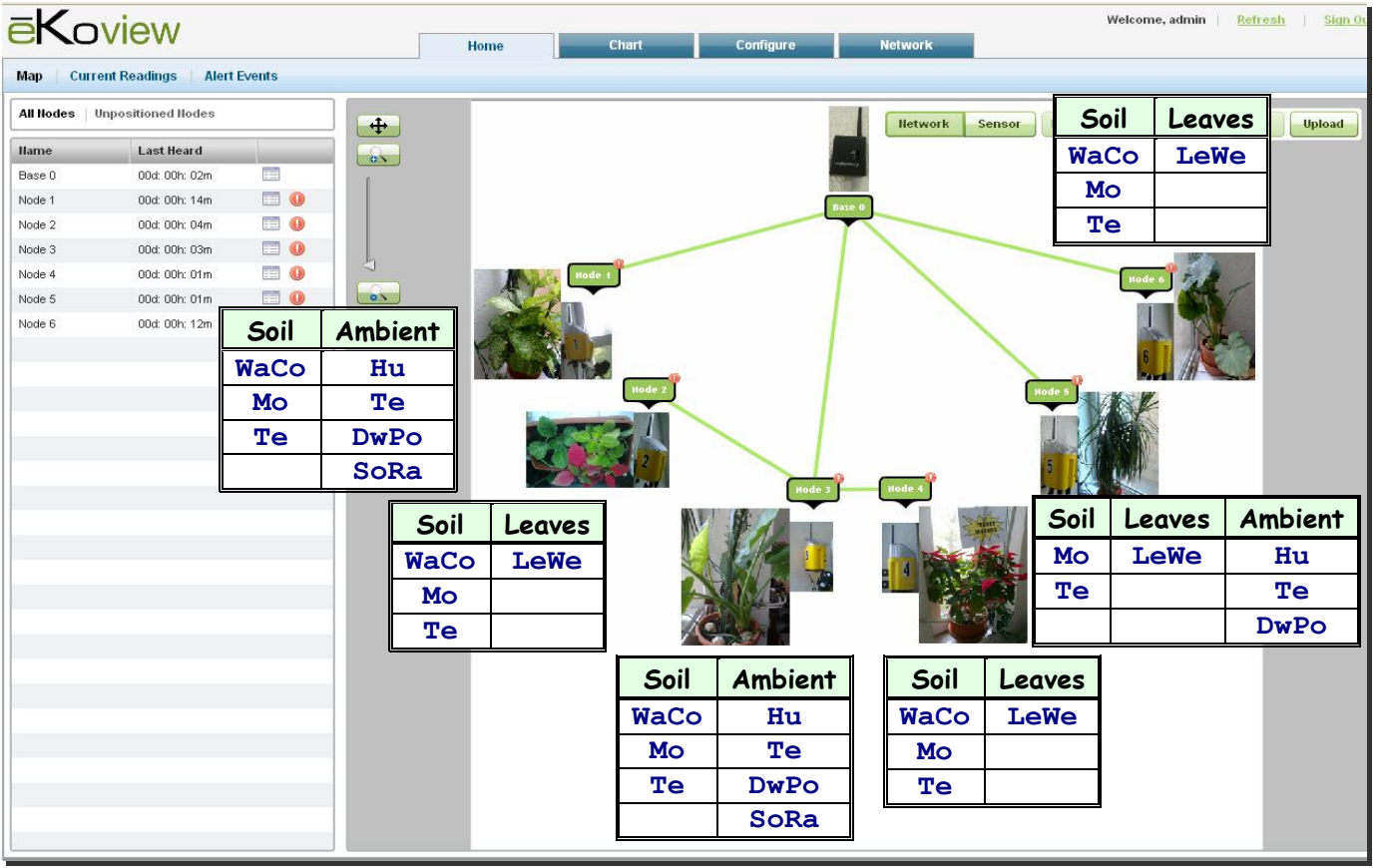

Fig. 3. Synoptic map of the monitored ecological parameters inside the greenhouse

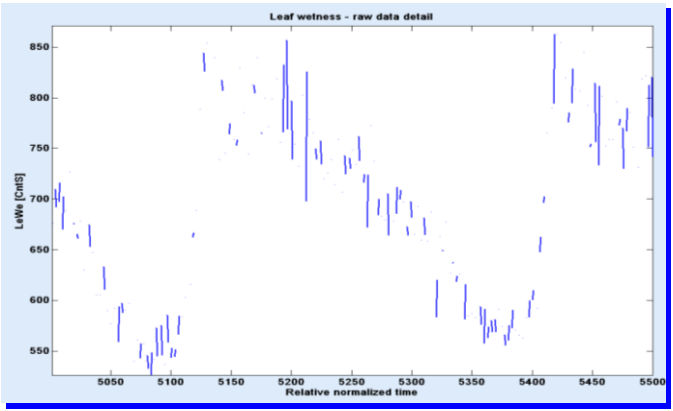

Fig. 4. Raw data for leaf wetness parameter

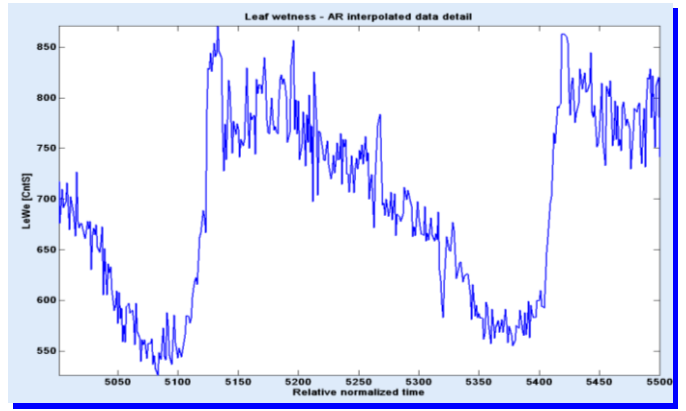

Fig. 6. AR interpolation of data for leaf wetness

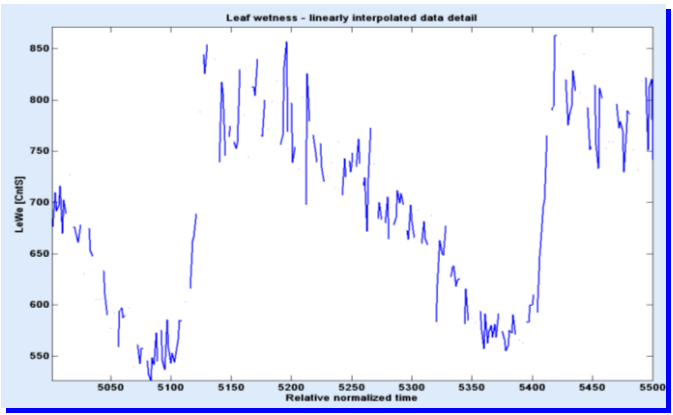

Fig. 5. Linear interpolated data for leaf wetness

There also occurs over-sampling phenomena, which means gathering much many samples than necessary. Here, an under-sampling technique is applied (for example, averaging). In our case, the data were averaged over 3-4 hours, especially for prediction, since the evolution of ecological phenomena is rather slow. Unlike before, it hardly happens that data contain important discrepancies (deviations) on short time intervals. These errors are eliminated by numerical filtering. One of the matching filters is second type Cebyshev (Proakis and Manolakis, 1996). For the ecological parameters this filter was indirectly applied with the aim of a refined delimitation between the deterministic and nondeterministic components of the prediction model.

## 4. SIMULATION RESULTS

The automatic irrigation application intended to improve the comfort and health of plants in the greenhouse, relatively to the situation of inappropriate watering. For the automatic control application, the parameters of interest are: soil moisture (Mo) and soil water content (WaCo). However, both parameters are correlated with soil temperature (Te). Therefore soil Te is one of the parameters to be predicted/monitored. Our simulations are focusing next on this parameter only (although in correlation with the other soil parameters).

The forecasting of greenhouse parameters is performed by means of PARMA, PARMAX, KARMA and FORWAVER predictors. A collection of 30 data blocks has been employed to predict various parameters. The data blocks resulted from combinations of soil or ambient parameters, as shown by the synoptic map of figure 3. The PARMAX predictor was the most employed since it has to be run several times for each channel. In order to reduce the simulation time, the EcoMonFor-F computer network was extended up to 16 PCs, including the laptop of EcoMonFor-M. The ecological phenomena usually act slowly. Therefore it is suitable to predict values every 3-4 hours. The simulation time for predictors varied between several minutes and several tens of hours, depending on their complexity, the number of analyzed ecological data and the modeling of stochastic component. Each one of the 30 data files is associated to 16 graphics for every acquisition channel, coming from all four predictors. There are 4 variations for a channel, which are bond to a predictor performance: the original data (time series) together with its optimal trend, the estimated white noise on measuring horizon; the predicted values and the prediction quality (Stefanoiu and Culita, 2010). Each predicted value has a trusting probability defined by the confidence tube. As the prediction instant goes away from the measuring horizon, the tube becomes larger and larger. This means the predicted values are less and less reliable.

In order to demonstrate the prediction performance of EcoMonFor, the soil Te from all 6 plants has been selected (as already stated). Figure 7 displays Te variations over the greenhouse, together with their best detected trends. When using PARMA or FORWAVER predictors, no correlations between soil Te and other parameters are considered. On the contrary, with PARMAX and KARMA, soil Te was predicted when considering correlations with soil moisture (Mo) and soil water content (WaCo) or leaf wetness (LeWe), as the chart of figure 3 already pointed.

Figures 8–13 reveal the prediction performance for soil Te, within each one of the 6 nodes. Best results of the 4 predictors (PARMA, PARMAX, KARMA and FORWAVER) are depicted, together with their corresponding *prediction quality* (PQ) values. (PQ was defined, for example, in (Stefanoiu and Culita, 2010)). The higher PQ the better. Although the predicted values are apparently very close to the real data, all variations were scaled in terms of trusting tube diameter (also drawn on all pictures).

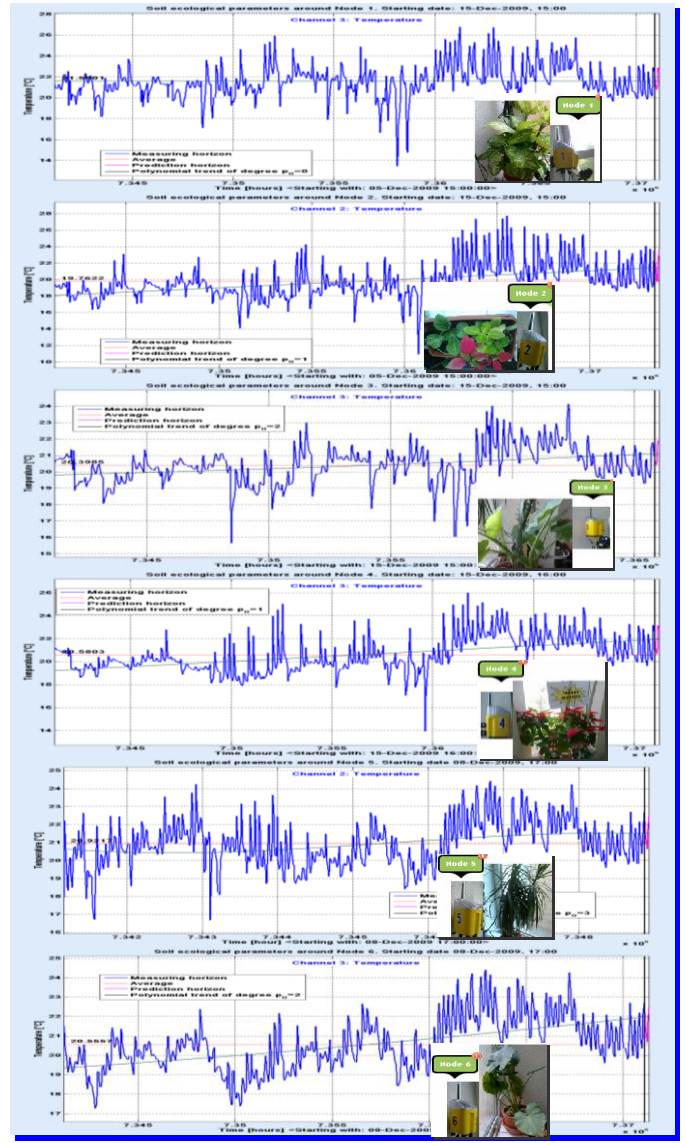

Fig. 7. Soil Te variations within the greenhouse.

So, the PQ values may take small values, just because the tube is too wide. As a general result, PARMA is never the best, but the fastest. However, its performance is fair, with a good trade-off between speed and accuracy, which allows assigning this predictor the bronze medal. For the silver medal, PARMAX is the righteous selection. Data on nodes 5 and 6 are best predicted with PARMAX. This time, correlations between soil Te and the other parameters helped the predictor to provide the best results. It must be noted however that PARMAX is by far the slowest predictor. The gold medallist is FORWAVER, with 4 best predicted values out of 6. However, like PARMA, this predictor is not accounting for correlations between parameters. Recall that FORWAVER is based on orthogonal wavelets from Daubechies class (Daubechies, 1988) and ARMA modelling of stochastic component (Stefanoiu et al., 2008).

A surprise, but a deceiving one, is made by KARMA, which performed much worst than expected (not only for Te, but for the other parameters as well). As one can easily notice the prediction results of KARMA are modest on all 6 channels.

A possible explanation resides in Kalman filter over-sensitivity to the variation of internal states number. Just removing or adding one single state can dramatically modify the predicted values outside as well as inside the measure horizon. The bronze-silver-gold classification is confirmed by all tests, with different greenhouse parameters (although, sometimes PARMAX is better than FORWAVER).

## 5. CONCLUSION

This article approached the problem of monitoring and forecasting of small greenhouse parameters. In subsidiary, the control system of greenhouse is also mentioned. The monitoring system (namely, EcoMonFor) integrates three user friendly interfaces eko-View, eko-Greenhouse and eko-Forecast, which are implemented on a mobile or fixed web computer. In order to model and predict the greenhouse evolution, the measured parameters are collected in data blocks depending on their correlation degree. Then, some preliminary operations (interpolation, numerical filtering) are applied for improving the quality of data. A comparative study between four predictors performance (PARMA, PARMAX, KARMA, FORWAVER) revealed that the best accuracy of prediction is achieved by PARMAX; next come PARMAX or PARMA, depending on the correlation degree between the monitored ecological parameters.

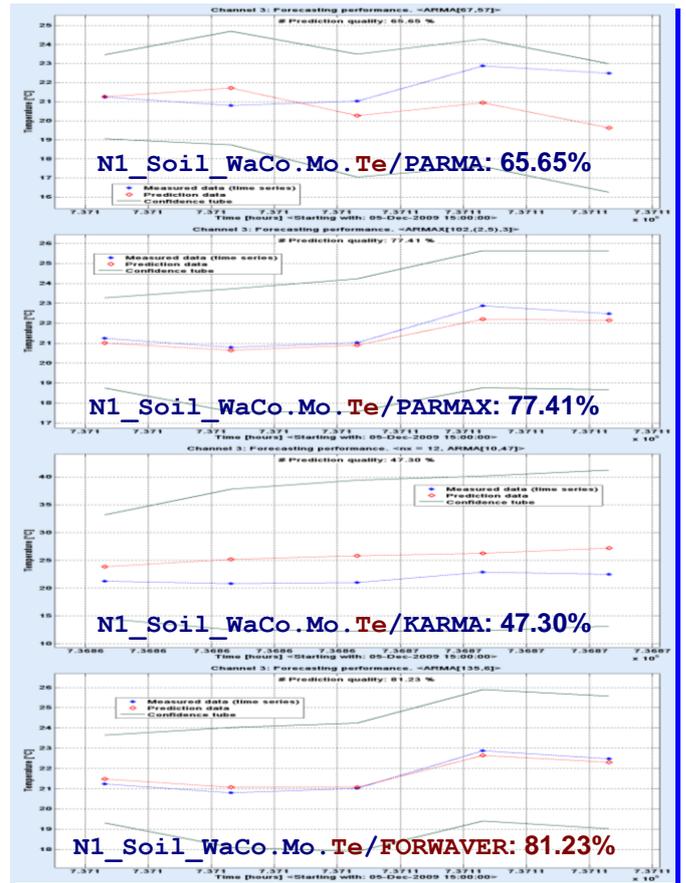

Fig. 8. Forecasting performance in node 1 (soil **Te**).

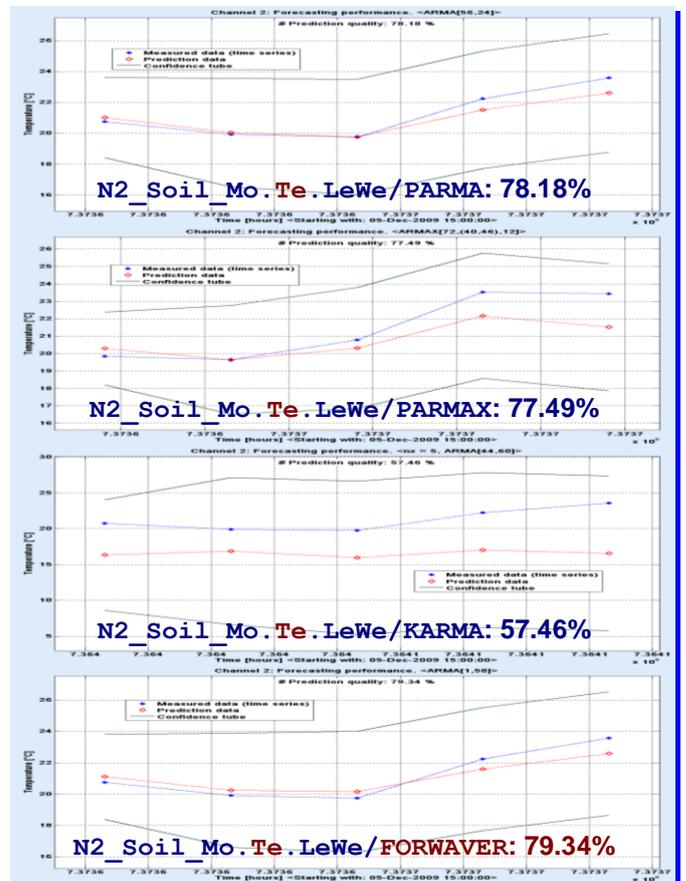

Fig. 9. Forecasting performance in node 2 (soil **Te**).

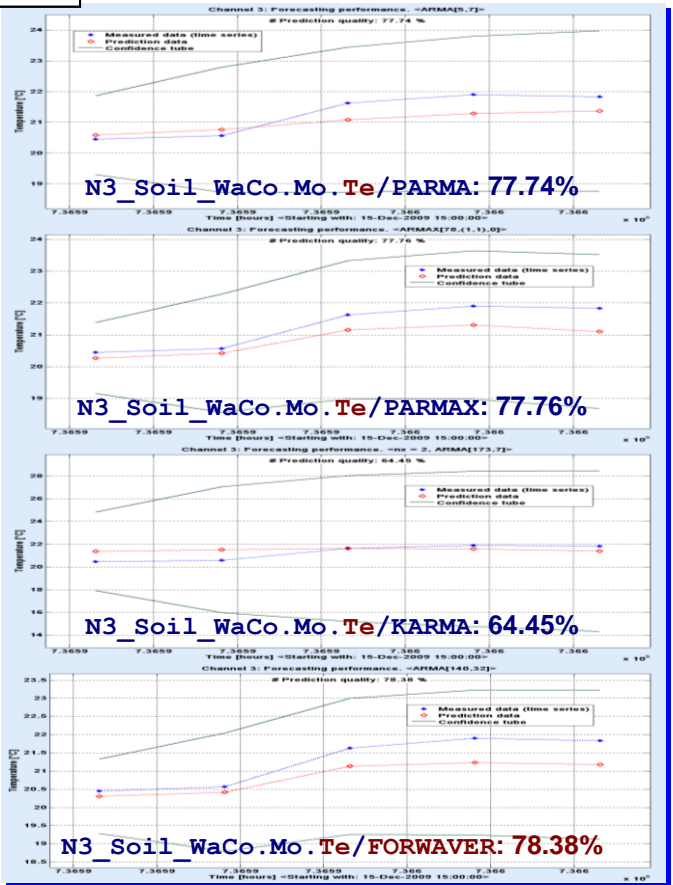

Fig. 10. Forecasting performance in node 3 (soil `Te`).

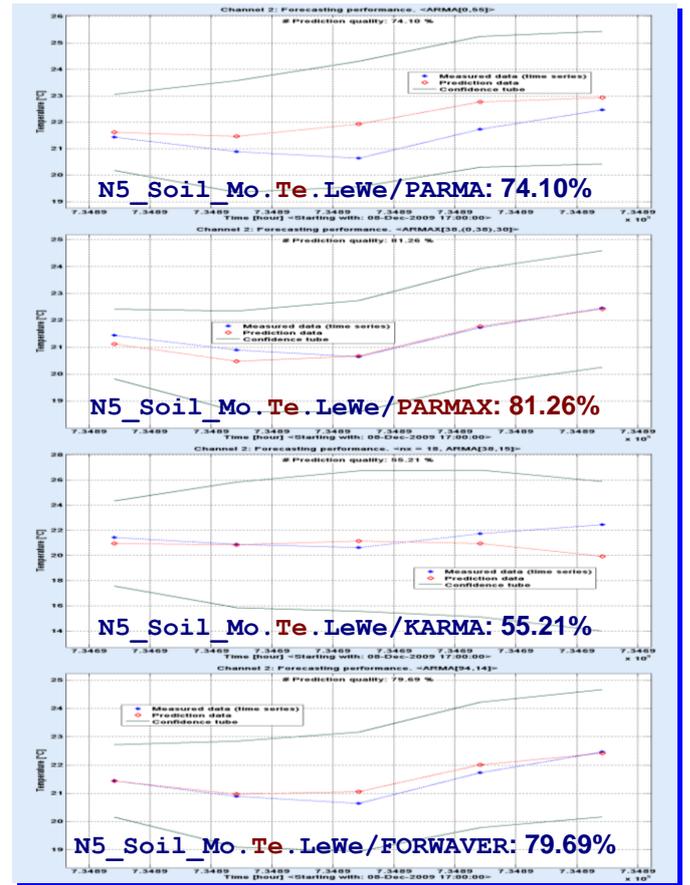

Fig. 12. Forecasting performance in node 5 (soil `Te`).

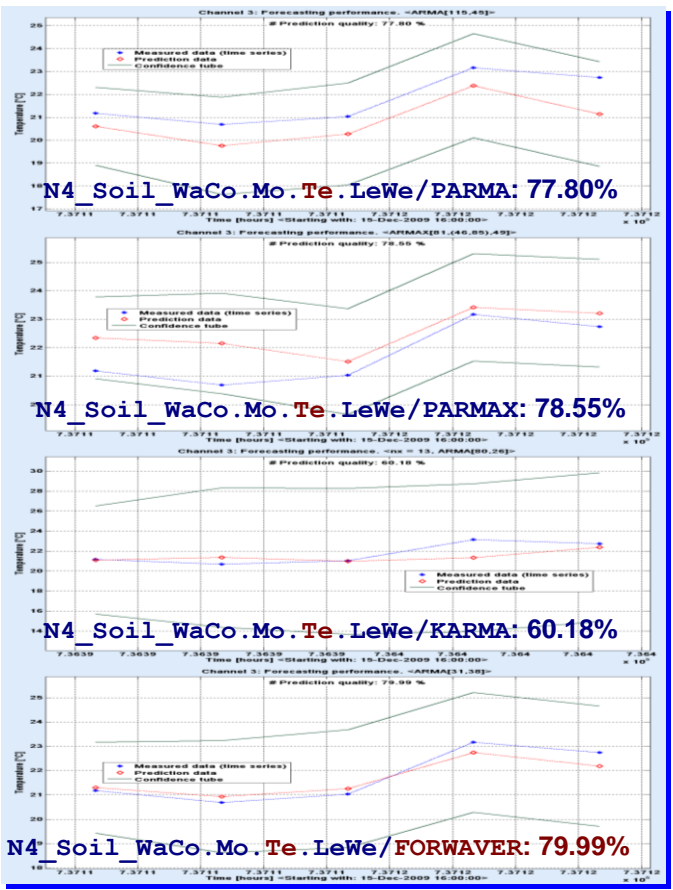

Fig. 11. Forecasting performance in node 4 (soil `Te`).

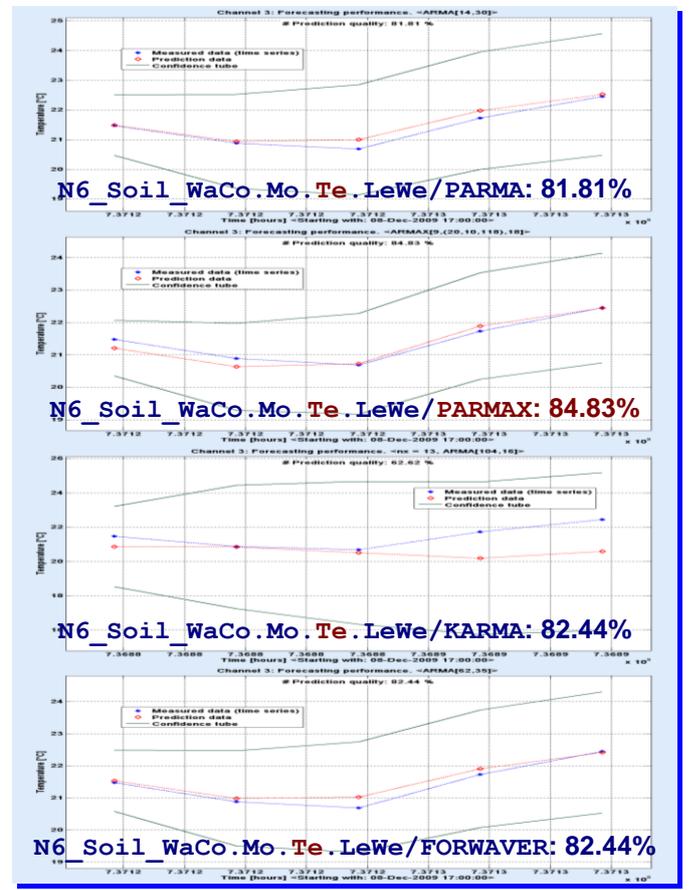

Fig. 13. Forecasting performance in node 6 (soil `Te`).